\documentclass[aps,prl,reprint,superscriptaddress,showpacs]{revtex4-1}
\usepackage{xspace}
\usepackage{graphicx}
\usepackage{siunitx}

\begin{document}

\newcommand{\msz}{$m_s=0$\xspace}
\newcommand{\mso}{$m_s=1$\xspace}
\newcommand{\mspmo}{$m_s=\pm1$\xspace}
\newcommand{\nvz}{NV$^0$\xspace}
\newcommand{\nvm}{NV$^-$\xspace}
\newcommand{\fidelity}{$\mathscr{F}_{C}$\xspace}
\newcommand{\mfid}{\mathscr{F}_{C}}
\newcommand{\tr}{$t_R$\xspace}
\newcommand{\mtr}{t_R}
\newcommand{\meanDm}{\langle D^-\rangle}
\newcommand{\meanDz}{\langle D^0\rangle}
\newcommand{\sigmaDm}{\sigma_{D^-}}
\newcommand{\sigmaDz}{\sigma_{D^0}}

\title{Efficient readout of a single spin state in diamond via spin-to-charge conversion} 

\author{B. J. Shields}
\author{Q. P. Unterreithmeier}
\author{N. P. de Leon}
\affiliation{Harvard University Physics Department}

\author{H. Park}
\affiliation{Harvard University Department of Chemistry \& Chemical Biology}
\affiliation{Harvard University Physics Department}

\author{M. D. Lukin}
\email{lukin@physics.harvard.edu}
\affiliation{Harvard University Physics Department}

\date{\today}

\begin{abstract}
Efficient readout of individual electronic spins associated with atom-like impurities in the solid state is  essential for applications in  quantum information processing and quantum metrology. We demonstrate a new method for efficient spin readout of nitrogen-vacancy (NV) centers in diamond. The method is based on conversion of the electronic spin state of the NV to a charge state distribution, followed by single-shot readout of the charge state.  Conversion is achieved through a spin-dependent photoionization process in diamond at room temperature.  Using NVs in nanofabricated diamond beams, we demonstrate that the resulting spin readout noise is within a factor of three of the spin projection noise level.  Applications of this technique for nanoscale magnetic sensing are discussed.
\end{abstract}


\maketitle

The negatively charged nitrogen-vacancy (NV) center in diamond is a solid state, atom-like impurity that combines a long lived spin-triplet ground state with an optical mechanism for both polarizing and reading out the electronic spin state at room temperature. These features make the NV center attractive for many applications such as nanoscale sensing\cite{Gopi2008,Jero2008,Kucsko2013kx,Toyli21052013} and quantum information processing\cite{Neumann:2010la,Dutt01062007,vanderSar2012}.  While the ability to optically detect the spin state at room temperature has enabled remarkable advances in diverse areas, this readout mechanism is not perfect.  Typically, single shot optical detection of quantum states in isolated atoms and atom-like systems requires a so-called cycling transition that can scatter many photons while returning to the original state.  Such cycling transitions exist at low temperature for the NV center, but at room temperature they cannot be selectively driven by laser excitation, due to phonon broadening.   Consequently, hundreds of repetitions are required to accurately distinguish between a spin prepared in \msz versus \mso. While single shot readout of the electronic spin has been observed, it is either slow (as in the case of repetitive readout involving nuclear ancilla\cite{Jiang09102009,PhilippNeumann07302010}) or requires cryogenic temperatures\cite{Robledo:2011sf}.

It is well known that the NV center can exist in several charge states.  In addition to \nvm, the neutral charge state (\nvz) has attracted recent interest for superresolution microscopy\cite{doi:10.1021/nl102156m,1367-2630-14-12-123002,PhysRevLett.106.157601,PhysRevLett.109.097404}.  Photoionization between the two charge states is well established\cite{1367-2630-15-1-013064, Manson20051705}.  However, previous studies of the charge state dynamics have focused on ionization timescales that are much longer than the internal dynamics of the \nvm energy levels, specifically the lifetime of the metastable singlet state.  Studies in this regime have established the charge state as a stable and high-contrast degree of freedom for fluorescence imaging, but have not explored the effect of spin on ionization.  In this Letter, we investigate photoionization on time scales relevant to the singlet state dynamics.  In this regime, we demonstrate a method for spin-to-charge conversion (SCC) that can be used for fast, efficient readout of the electronic spin state of the NV center.

\begin{figure}
\includegraphics{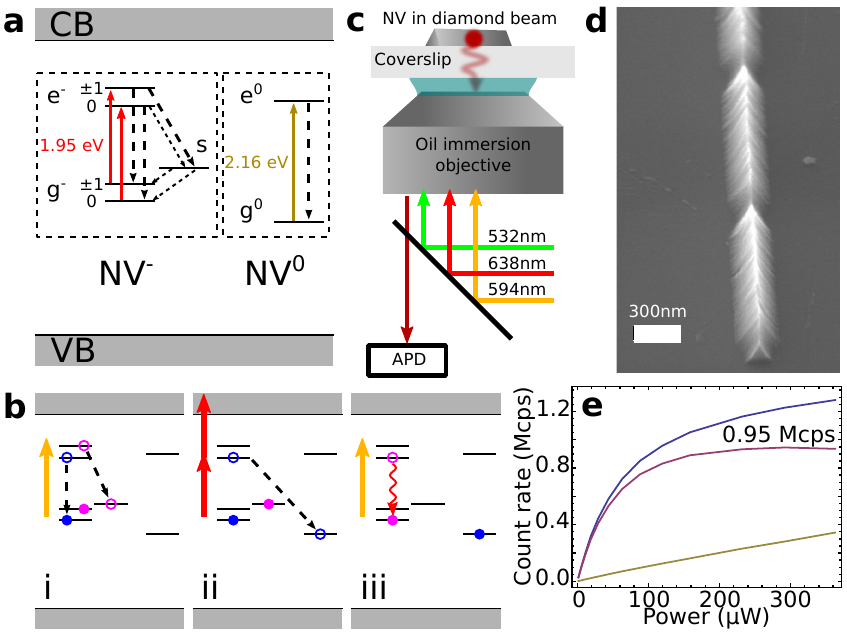}
\caption{\label{fig:SetupFig}SCC measurement idea. (a) Level diagram for \nvm and \nvz, indicating triplet ground ($g^-$) and excited ($e^-$), and metastable singlet ($s$) states of \nvm, and ground ($g^0$) and excited ($e^0$) states of \nvz. (b) SCC measurement process for an initial state of \msz (blue) or \mso (pink).  (i) A 594-nm pulse either shelves into the singlet state (\mso) or cycles (\msz). (ii) A 638-nm pulse rapidly ionizes population from the \nvm triplet states to \nvz. (iii) Single-shot charge state measurement with 594-nm light. (c) Setup for high collection efficiency from diamond nanobeams.  Fluorescence is collected with an oil immersion microscope and imaged onto a multimode fiber. (d) SEM micrograph of a diamond nanobeam transferred to silicon, imaged at \SI{60}{\degree} tilt.  (e) Saturation fluorescence measurement for an NV in a diamond nanobeam.  Total fluorescence (blue), background from glass (gold), NV signal (red). The maximum count rate for cw 532-nm illumination is 0.945 Mcps after background subtraction.}
\end{figure}

The key component of the SCC method is a two-step pulse sequence that rapidly transfers the spin state of \nvm to a charge distribution, as illustrated in Fig.~\ref{fig:SetupFig}.  This mechanism is related to the well-established technique for optically detected magnetic resonance (ODMR)\cite{Gruber27061997}, in that it takes advantage of the spin-dependent shelving process to the metastable singlet state.  Specifically, we utilize the fact that, upon 594-nm excitation, the $m_s=\pm1$ states of \nvm can be optically shelved into a metastable singlet manifold via an intersystem crossing, while the $m_s=0$ state cycles within the manifold of triplet ground and excited states.  Subsequently,  the \nvm triplet excited state can be ionized using a second intense pulse of 638-nm light, but the \nvm singlet manifold cannot be excited back to the triplet excited state by either the 594-nm or 638-nm light, and hence is protected from ionization. Thus, \nvm in the \msz state will be ionized to \nvz upon two-pulse excitation, whereas \nvm in the \mspmo state will remain mostly as \nvm.  Single-shot charge-state detection then provides a sensitive measurement of the electron spin state.  The stability and spectral contrast of the charge states minimizes the contribution of photon shot noise, so that the measurement is instead limited by the SCC efficiency.  As a result, the readout noise is dramatically reduced, to a limit of $\sim$2.76 times the spin projection noise level.

For our measurements we use naturally occurring NVs in type IIa chemical vapor deposition grown diamond (Element6, 1 ppm N concentration).  To enhance the photon collection efficiency, we carve the diamond into nanobeams and transfer them to a glass coverslip for imaging in an oil-immersion confocal microscope (Fig.~\ref{fig:SetupFig}c).  We fabricate the nanobeams with an angled reactive ion etching technique\cite{doi:10.1021/nl302541e} that yields triangular cross-section waveguides with a width of \SI{300}{\nano\meter} and a length of \SI{20}{\micro\meter}, suspended above the diamond substrate.  In the same step, we etch notches (\SI{50}{\nm} depth) every \SI{2}{\um} along the beam, to scatter waveguided light.  Using a \SI{500}{\nano\meter} radius tungsten probe tip mounted on a 3-axis piezostage, we detach the beams from the diamond, place them on the coverslip, and orient them so that the smooth, unetched diamond surface contacts the glass.

To address the NV optically, we illuminate it through a microscope objective (Nikon, NA=1.49) with laser light at 532-, 594-, and 638-nm wavelengths (Fig.~\ref{fig:SetupFig}c), which serve to pump the charge state into \nvm, drive \nvm to the triplet excited state, and ionize from the \nvm triplet manifold to \nvz, respectively.  The timing and intensity of each laser is controlled by an acousto optic modulator (AOM).  We collect fluorescence from the NV through the same objective and image it onto a multimode fiber.

Recent work on high collection efficiency with immersion imaging systems relied upon the placement of an emitter in a low-index layer on top of a high-index substrate\cite{G2011uq,Maletinsky2014}.  Due to the high refractive index of diamond ($n_{diamond}=2.4$), however, obtaining a substrate of higher index is difficult. Instead, we use the subwavelength dimension of the nanobeams to avoid total internal reflection at the diamond surface, so that the NV fluorescence is efficiently coupled to radiative modes in the glass.   In this way we observe a maximum count rate of 0.945 million counts per second (cps) under cw 532-nm illumination (Fig.~\ref{fig:SetupFig}e).  To manipulate the \nvm electron spin sublevels, we align the magnetic field from a permanent magnet with the NV axis, splitting \mspmo. A copper wire (\SI{25}{\micro\meter} diameter) adjacent to the beams delivers a \SI{2.917}{\giga\hertz} microwave field to drive transitions between \msz and \mso.

\begin{figure}
\includegraphics{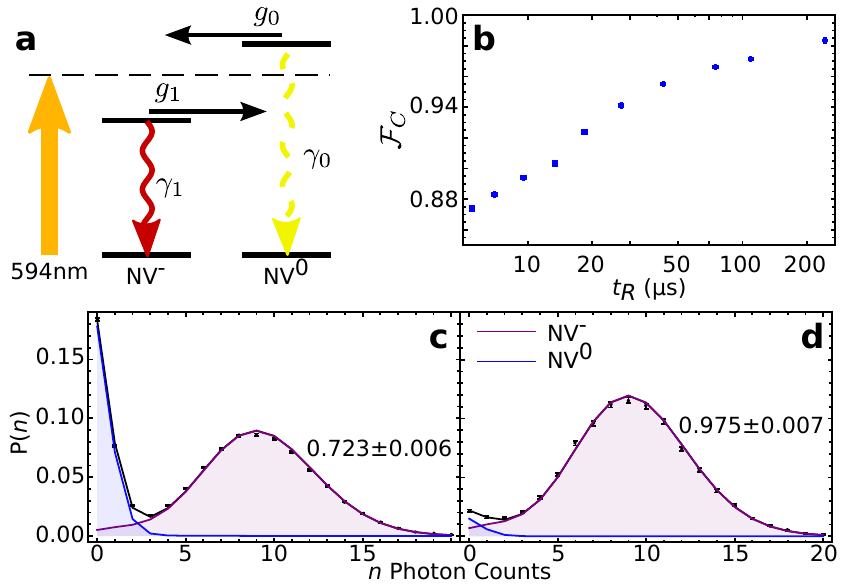}
\caption{\label{fig:ChargeReadoutFig}NV charge state initialization and readout. (a) Level diagram for the charge state readout process.  594-nm light efficiently excites \nvm, while weakly exciting \nvz, resulting in photon count rates $\gamma_1$, $\gamma_0$, respectively.  Ionization occurs from the excited states of each charge configuration at rates $g_1$, $g_0$. (b) Optimized charge state readout fidelity $\mathcal{F}_C$ as a function of readout time, determined by measuring $\gamma_1$, $\gamma_0$, $g_1$, and $g_0$ at various illumination intensities.  (c) Fidelity of charge state initialization.  A pump-probe sequence initializes into \nvm (see text), and a \SI{2.24}{\ms} readout (594-nm, \SI{820}{\nano\watt}) following the probe verifies the charge state.  The readout photon number distribution for 100,000 iterations (left) indicates an initialization fidelity of $0.723\pm0.006$.  Conditioning on the observation of one or more probe photons (right), the fidelity increases to $0.975\pm0.007$.  The black line is a fit to the full \SI{2.24}{\ms} readout, and the blue and purple lines indicate the fitted \nvz and \nvm contributions, respectively.}
\end{figure}

Central to our spin readout process is a mechanism for high-fidelity measurement of the NV charge state\cite{1367-2630-15-1-013064}. This measurement utilizes the different excitation and emission spectra for \nvm and \nvz, allowing for efficient spectral discrimination.  A low power of 594-nm light efficiently excites the \nvm sideband, but only weakly excites \nvz (Fig.~\ref{fig:ChargeReadoutFig}a).  A 655-nm longpass filter is used in the collection path to eliminate any residual \nvz fluorescence.  In this way, \nvm can be made 20-30 times brighter than \nvz (depending on laser intensity\cite{sup}), resulting in a high contrast measurement.

Laser illumination also causes the NV to jump between charge states\cite{1367-2630-15-1-013064}.  The NV first absorbs one photon and then, while in an excited configuration, absorbs a second photon, either exciting an electron to the conduction band to ionize \nvm to \nvz, or recapturing an electron from the valence band to convert \nvz to \nvm.  Thus, at low power, the ionization and recapture rates, $g_1$ and $g_0$, respectively, obey a quadratic power dependence, whereas the \nvz and \nvm photon count rates, $\gamma_0$ and $\gamma_1$, obey a linear power dependence\cite{1367-2630-15-1-013064}.  Consequently, the illumination power and integration time of the measurement can be adjusted to allow faster readout at the expense of lower readout fidelity.

To characterize the charge state readout fidelity, $\mathcal{F}_C$, of our setup, we measure the four rates, $g_{0,1}$, $\gamma_{0,1}$, under cw 594-nm illumination, for powers ranging from \SI{0.875}{\micro\watt} to \SI{15}{\micro\watt}.  At each power, we record the number of photons detected in a time window, $t \sim 1/g_1$ (so that the resulting photon number statistics are sensitive to the ionization rates).  We then fit the photon number distribution for 100,000 time windows with a model for the charge state dynamics, to obtain the four rates at each power\cite{sup}.  From the measured rates at a power $P$, we calculate the optimal readout time $t_R$ to maximize $\mathcal{F}_C(P)$ (Fig.~\ref{fig:ChargeReadoutFig}b).  We obtain high fidelity ($\mathcal{F}_C \sim 0.9$) even for readout times as short as \SI{10}{\micro\second}.

A similar measurement scheme can be used to rapidly initialize the NV into \nvm.  To do so, we apply a short, high power pump pulse of 532-nm light (\SI{150}{\nano\second} at \SI{300}{\micro\watt}), and then measure the charge state with a short probe pulse of 594-nm light ($t_{\mathrm{probe}}=\SI{900}{\nano\second}$ at \SI{11}{\micro\watt}).  In this regime, $g_1t_{\mathrm{probe}}\ll 1$, so that ionization is unlikely and detection of 1 or more photons verifies that the final charge state is \nvm.  Failed verification attempts can be discarded.

To verify our initialization fidelity, we perform a pump-probe combination followed by charge state readout at low power ($t_R=\SI{2.24}{\ms}$ at \SI{820}{\nano\watt}).  The readout time is longer than optimal in order to obtain an accurate fit of the populations.  The photon number distribution for 100,000 measurements is shown in Fig.~\ref{fig:ChargeReadoutFig}c,d, where we plot the distribution for the first \SI{240}{\micro\second} of readout for clarity.  Figure~2c shows results for all probe outcomes, indicating an initialization fidelity of $0.723 \pm 0.006$.  Figure~2d shows the distribution conditioned on the detection of one or more probe photons, for which the initialization fidelity increases to $0.975 \pm 0.007$.  For these pump-probe conditions, a single initialization step ``succeeds'' (detects one or more probe photons) with probability $p_s = 0.216\pm 0.001$.

We next demonstrate spin dependent control of the ionization dynamics, allowing for efficient conversion from the \nvm electron spin state to a charge state distribution.  We first initialize into \nvm and prepare the spin into either \mso or \msz, and then apply a short, intense pulse of 594-nm light (\SI{145}{\micro\watt}) that drives \nvm into its triplet excited state (Fig.\ref{fig:SetupFig}b(i)).  Depending on initial spin state preparation, the triplet excited state either decays into the singlet state via an intersystem crossing (in the case of \mso), or relaxes back to the triplet ground state (in the case of \msz).  Following the first 594-nm pulse, we immediately apply a short, high power pulse of 638-nm light (\SI{22.5}{\milli\watt}), to rapidly ionize any population remaining in the triplet manifold (Fig.~\ref{fig:SetupFig}b(ii)).  This pulse does not excite the singlet manifold of \nvm, leading to spin dependent ionization, and thus spin-to-charge conversion.  Finally, we measure the charge state of the NV (Fig.~\ref{fig:SetupFig}b(iii)).

\begin{figure}
\includegraphics{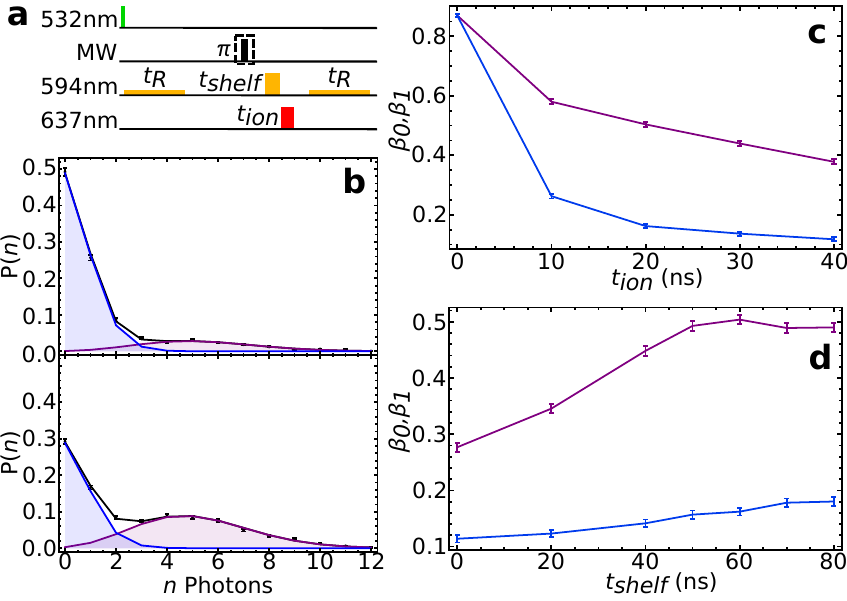}
\caption{\label{fig:SCCFig}Spin to charge conversion.  (a) Pulse sequence consisting of initialization with \SI{40}{\nano\second}, \SI{300}{\micro\watt} pump pulse and \SI{500}{\micro\second}, \SI{500}{\nano\watt} probe pulse, followed by microwave manipulation to prepare in either \msz or \mso, two-pulse SCC for duration $t_{shelf}$ and $t_{ion}$, and finally charge readout for \SI{500}{\micro\second} at \SI{500}{\nano\watt}.  (b) Photon number distributions for $t_{shelf}=\SI{60}{\nano\second}$ and $t_{ion}=\SI{20}{\nano\second}$.  An initial state of \msz ionizes to \nvz (top, \nvm population = $0.162 \pm 0.007$) while an initial state of \mso is shelved into the singlet state and protected from ionization (bottom, \nvm population = $0.504 \pm 0.009$).  (c) Final \nvm population for $t_{ion}$ ranging from \SI{0}{\nano\second} to \SI{40}{\nano\second} ($t_{shelf}=\SI{60}{\nano\second}$). (c) Final \nvm population for $t_{shelf}$ ranging from \SI{0}{\nano\second} to \SI{80}{\nano\second} ($t_{ion}=\SI{20}{\nano\second}$), showing the dynamics of the shelving process.}
\end{figure}

The resulting photon number distributions are shown in Fig.~\ref{fig:SCCFig}b, for an initial spin state of \msz (top) and \mso (bottom), where we use a shelving pulse duration $t_{shelf}=\SI{60}{\ns}$ and an ionization pulse duration $t_{ion}=\SI{20}{\ns}$.  From a fit to the measured photon number distributions, we determine the average population in \nvm at the end of the SCC step.  For an initial state of \msz or \mso, we label the average final \nvm population $\beta_0$ or $\beta_1$.  The contrast between $\beta_0$ and $\beta_1$ characterizes the efficiency of the SCC mechanism.  To optimize the SCC efficiency we sweep both $t_{shelf}$ and $t_{ion}$ over a range of times, as shown in Fig.~\ref{fig:SCCFig}c,d.  In Fig.~\ref{fig:SCCFig}c, $t_{shelf}$ is fixed at 60ns and we sweep $t_{ion}$.  For each $t_{ion}$, we measure the photon number distributions as in Fig.~\ref{fig:SCCFig}b to find $\beta_{0,1}(t_{ion})$.  Similarly, in Fig.~\ref{fig:SCCFig}d, we fix $t_{ion}=\SI{20}{\nano\second}$ and sweep $t_{shelf}$.  As $t_{shelf}$ is increased, the \mso population is transferred to the singlet state and protected from ionization, resulting in a maximum for $\beta_1(t_{shelf})$ at $t_{shelf}=\SI{60}{\nano\second}$.

To quantify the performance of the SCC mechanism for \nvm electronic spin readout, we consider its applications for  magnetometry\cite{Taylor:2008nx}.  We consider a magnetometry sequence based on a Hahn echo\cite{PhysRev.80.580}, and compare the readout noise for the SCC scheme with the conventional ODMR readout mechanism.  In both cases the magnetic field sensitivity is:
\begin{equation}\label{eq:magnetometersensitivity}
	\eta = \frac{\pi\hbar}{2g\mu_B} \times \sigma_R \times \sqrt{\frac{\tau + t_I + t_R}{\tau^2}},
\end{equation}
where $g$ is the electron gyromagnetic ratio, $\mu_B$ is the Bohr magneton, $\tau$ is the Hahn echo time, $t_I$ is the initialization time, and $t_R$ is the spin readout time.  $\sigma_R$ is a measure of the spin readout noise for a single measurement, normalized so that $\sigma_R=1$ for a perfect measurement (i.e. limited by only the fundamental quantum spin projection noise).  In the case of SCC readout, both $\sigma_R$ and the measurement duty cycle depend on $t_R$, so the optimal readout conditions will vary depending on $\tau$.

In the conventional spin readout scheme, the NV is prepared into \nvm, the Hahn echo is applied for a time $\tau$ and the spin is read out with a short excitation pulse (typically $\sim$ \SI{200}{\nano\second} of 532-nm light), during which time an average number of photons $\alpha_0$ or $\alpha_1$ is counted when the NV is projected into \msz or \mso, respectively.  The two sources of noise in this case are spin projection noise and photon shot noise, and the overall spin readout noise is\cite{Taylor:2008nx}:
\begin{equation}
	\sigma_R = \sqrt{1+\frac{2(\alpha_0+\alpha_1)}{(\alpha_0-\alpha_1)^2}}.
\end{equation}
For a bulk diamond sample, typical photon collection efficiencies result in a best-case value of $\sigma_R\sim 20$\cite{Balasubramanian:2009bf}.  With the enhanced collection efficiency from the diamond nanobeam geometry, we observe $\alpha_0=0.238\pm 0.001$ and $\alpha_1=0.154\pm 0.002$, resulting in $\sigma_R=10.6\pm 0.3$.  In both cases, photon shot noise is by far the dominant source of noise.

In the case of SCC readout, the final charge state is measured by counting photons and assigning the result to \nvz or \nvm based on a threshold photon number.  The probability of measuring \nvm in this way is $\tilde{\beta}_0$ or $\tilde{\beta}_1$ for an initial spin state of \msz or \mso, and the spin readout noise is then given by:
\begin{equation}\label{eq:SCCnoise}
	\sigma_R^{SCC} = \sqrt{\frac{(\tilde{\beta}_0+\tilde{\beta}_1)(2-\tilde{\beta}_0-\tilde{\beta}_1)}{(\tilde{\beta}_0-\tilde{\beta}_1)^2}}.
\end{equation}
In the limit of perfect charge readout, $\tilde{\beta}_{0,1}$ approach the true charge state population values $\beta_{0,1}$.  For the optimized SCC process in Fig.~\ref{fig:SCCFig}b, this corresponds to $\sigma_{R,\mathrm{min}}^{SCC}=2.76\pm 0.09$.  Note that this includes the effects of imperfect initial spin polarization (measured to be $92\pm1$\% in our system\cite{sup}) and imperfect charge initialization.
  
\begin{figure}
\includegraphics{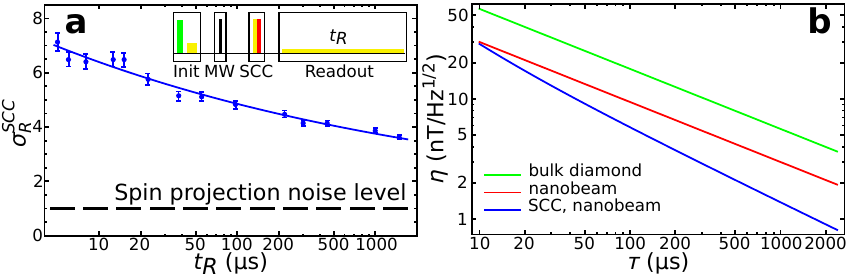}
\caption{\label{fig:SigmaRVsTime}Time dependence and magnetometer sensitivity.  (a) Measurement of $\sigma_R^{SCC}(t_R)$ (blue points) fit by a power law (solid line).  The inset shows the pulse sequence used, consisting of initialization (\SI{150}{\nano\second}, \SI{300}{\micro\watt}, 532-nm pump, \SI{900}{\nano\second}, \SI{11}{\micro\watt} 594-nm probe), microwave pulse to prepare in \msz or \mso, SCC sequence (\SI{50}{\nano\second}, \SI{135}{\micro\watt}, 594-nm shelving pulse followed by \SI{30}{\nano\second}, \SI{7.1}{\milli\watt}, 638-nm ionization pulse), and readout.  (b) Based on the fit in (a), we directly calculate the magnetometer sensitivity (Eq.~\ref{eq:magnetometersensitivity}) as a function of spin coherence time for the SCC readout (blue curve). The sensitivity for conventional readout in bulk diamond ($\sigma_R=20$, green curve) and diamond nanobeams ($\sigma_R=10.6$, red curve) is shown for comparison.}
\end{figure}

To evaluate the practical utility of SCC readout for magnetometry, we measured $\sigma_R^{SCC}(t_R)$ using the pulse sequence shown in Fig.~\ref{fig:SigmaRVsTime}a(inset), with the fast initialization scheme described above ($t_I=\SI{6.5}{\micro\second}$), over a range of values for $t_R$.  We optimized the readout power and threshold photon number for each $t_R$ so as to minimize $\sigma_R^{SCC}(t_R)$.  The results are shown in Fig.~\ref{fig:SigmaRVsTime}a.  For short $t_R\sim \SI{5}{\micro\second}$, $\sigma_R^{SCC}(t_R)$ provides a modest improvement over the conventional readout scheme. For longer readout times, the contribution from photon shot noise due to imperfect charge readout diminishes, and the noise improves by a factor of 3 over conventional readout.  

With the measurement of $\sigma_R^{SCC}(t_R)$, the magnetometer sensitivity can now be directly estimated from Eq.~\ref{eq:magnetometersensitivity}, as shown in Fig.~\ref{fig:SigmaRVsTime}b.  For the spin coherence times measured in our nanobeams  (\SI{200}{\us}\cite{sup}), we estimate a sensitivity of \SI{4}{\nano\tesla\per\hertz^{1/2}}, while for coherence times in the range of \SI{2}{\ms}, demonstrated in $^{12}$C isotopically pure diamond\cite{Balasubramanian:2009bf}, the sensitivity will be \SI{900}{\pico\tesla\per\hertz^{1/2}}.

Before concluding, we note that several improvements to the SCC method may be possible.  We expect $\sigma_R^{SCC}(t_R)$ to approach $\sigma_{R,\mathrm{min}}^{SCC}$ for long $t_R$, as photon shot noise becomes negligible.  However, the measured values for are somewhat higher.  We believe this is due to the pump duty cycle employed for fast initialization, which may have some effect on the ionization dynamics that is not fully described by our model.  Additionally, the limiting value, $\sigma_{R,\mathrm{min}}^{SCC}$, is set by the internal dynamics of \nvm, and by the photoionization cross section, which is material dependent.  For instance, it is known that the photoionization behavior in diamond with high defect density can be very different from that observed here, with the charge state being much less stable\cite{Manson20051705}. Therefore, it may be possible to obtain more favorable conditions for ionization dynamics, and thereby a lower value for $\sigma_{R,\mathrm{min}}^{SCC}$, by controlling the defect density in the crystal.  

To summarize, we have studied the ionization dynamics of the NV center on timescales commensurate with the internal spin dynamical processes of the \nvm charge state.  In particular, we have demonstrated a spin-dependent ionization process that maps the spin state of \nvm onto a charge distribution between \nvm and \nvz.  This mechanism provides a significant improvement in the spin readout noise of a single measurement shot, to a limit of $\sim$2.76 times the spin projection noise level.  This directly results in improved single-spin magnetometer sensitivity.  In addition to applications in nanoscale sensing, the selective ionization of the \nvm triplet manifold can be used to extend ionization-based studies of NV spectroscopy\cite{1367-2630-15-1-013064}.

\begin{acknowledgments}
We thank A.~ Gali, A.~Trifonov, S. Kolkowitz, A.~Sipahigil, T.~Tiecke, Y.~Chu, and A.~Zibrov for helpful discussions and experimental support.  This work was performed in part at the Harvard Center for Nanoscale Systems.  We acknowledge support from the NSF, CUA, HQOC, DARPA QuASAR program, ARO MURI, AFOSR MURI, and the Moore Foundation.  \end{acknowledgments}

\bibliography{References}

\begin{thebibliography}{26}%
\makeatletter
\providecommand \@ifxundefined [1]{%
 \@ifx{#1\undefined}
}%
\providecommand \@ifnum [1]{%
 \ifnum #1\expandafter \@firstoftwo
 \else \expandafter \@secondoftwo
 \fi
}%
\providecommand \@ifx [1]{%
 \ifx #1\expandafter \@firstoftwo
 \else \expandafter \@secondoftwo
 \fi
}%
\providecommand \natexlab [1]{#1}%
\providecommand \enquote  [1]{``#1''}%
\providecommand \bibnamefont  [1]{#1}%
\providecommand \bibfnamefont [1]{#1}%
\providecommand \citenamefont [1]{#1}%
\providecommand \href@noop [0]{\@secondoftwo}%
\providecommand \href [0]{\begingroup \@sanitize@url \@href}%
\providecommand \@href[1]{\@@startlink{#1}\@@href}%
\providecommand \@@href[1]{\endgroup#1\@@endlink}%
\providecommand \@sanitize@url [0]{\catcode `\\12\catcode `\$12\catcode
  `\&12\catcode `\#12\catcode `\^12\catcode `\_12\catcode `\%12\relax}%
\providecommand \@@startlink[1]{}%
\providecommand \@@endlink[0]{}%
\providecommand \url  [0]{\begingroup\@sanitize@url \@url }%
\providecommand \@url [1]{\endgroup\@href {#1}{\urlprefix }}%
\providecommand \urlprefix  [0]{URL }%
\providecommand \Eprint [0]{\href }%
\providecommand \doibase [0]{http://dx.doi.org/}%
\providecommand \selectlanguage [0]{\@gobble}%
\providecommand \bibinfo  [0]{\@secondoftwo}%
\providecommand \bibfield  [0]{\@secondoftwo}%
\providecommand \translation [1]{[#1]}%
\providecommand \BibitemOpen [0]{}%
\providecommand \bibitemStop [0]{}%
\providecommand \bibitemNoStop [0]{.\EOS\space}%
\providecommand \EOS [0]{\spacefactor3000\relax}%
\providecommand \BibitemShut  [1]{\csname bibitem#1\endcsname}%
\let\auto@bib@innerbib\@empty
\bibitem [{\citenamefont {Balasubramanian}\ \emph {et~al.}(2008)\citenamefont
  {Balasubramanian}, \citenamefont {Chan}, \citenamefont {Kolesov},
  \citenamefont {Al-Hmoud}, \citenamefont {Tisler}, \citenamefont {Shin},
  \citenamefont {Kim}, \citenamefont {Wojcik}, \citenamefont {Hemmer},
  \citenamefont {Krueger}, \citenamefont {Hanke}, \citenamefont
  {Leitenstorfer}, \citenamefont {Bratschitsch}, \citenamefont {Jelezko},\ and\
  \citenamefont {Wrachtrup}}]{Gopi2008}%
  \BibitemOpen
  \bibfield  {author} {\bibinfo {author} {\bibfnamefont {G.}~\bibnamefont
  {Balasubramanian}}, \bibinfo {author} {\bibfnamefont {I.~Y.}\ \bibnamefont
  {Chan}}, \bibinfo {author} {\bibfnamefont {R.}~\bibnamefont {Kolesov}},
  \bibinfo {author} {\bibfnamefont {M.}~\bibnamefont {Al-Hmoud}}, \bibinfo
  {author} {\bibfnamefont {J.}~\bibnamefont {Tisler}}, \bibinfo {author}
  {\bibfnamefont {C.}~\bibnamefont {Shin}}, \bibinfo {author} {\bibfnamefont
  {C.}~\bibnamefont {Kim}}, \bibinfo {author} {\bibfnamefont {A.}~\bibnamefont
  {Wojcik}}, \bibinfo {author} {\bibfnamefont {P.~R.}\ \bibnamefont {Hemmer}},
  \bibinfo {author} {\bibfnamefont {A.}~\bibnamefont {Krueger}}, \bibinfo
  {author} {\bibfnamefont {T.}~\bibnamefont {Hanke}}, \bibinfo {author}
  {\bibfnamefont {A.}~\bibnamefont {Leitenstorfer}}, \bibinfo {author}
  {\bibfnamefont {R.}~\bibnamefont {Bratschitsch}}, \bibinfo {author}
  {\bibfnamefont {F.}~\bibnamefont {Jelezko}}, \ and\ \bibinfo {author}
  {\bibfnamefont {J.}~\bibnamefont {Wrachtrup}},\ }\href
  {http://dx.doi.org/10.1038/nature07278} {\bibfield  {journal} {\bibinfo
  {journal} {Nature}\ }\textbf {\bibinfo {volume} {455}},\ \bibinfo {pages}
  {648} (\bibinfo {year} {2008})}\BibitemShut {NoStop}%
\bibitem [{\citenamefont {Maze}\ \emph {et~al.}(2008)\citenamefont {Maze},
  \citenamefont {Stanwix}, \citenamefont {Hodges}, \citenamefont {Hong},
  \citenamefont {Taylor}, \citenamefont {Cappellaro}, \citenamefont {Jiang},
  \citenamefont {Dutt}, \citenamefont {Togan}, \citenamefont {Zibrov},
  \citenamefont {Yacoby}, \citenamefont {Walsworth},\ and\ \citenamefont
  {Lukin}}]{Jero2008}%
  \BibitemOpen
  \bibfield  {author} {\bibinfo {author} {\bibfnamefont {J.~R.}\ \bibnamefont
  {Maze}}, \bibinfo {author} {\bibfnamefont {P.~L.}\ \bibnamefont {Stanwix}},
  \bibinfo {author} {\bibfnamefont {J.~S.}\ \bibnamefont {Hodges}}, \bibinfo
  {author} {\bibfnamefont {S.}~\bibnamefont {Hong}}, \bibinfo {author}
  {\bibfnamefont {J.~M.}\ \bibnamefont {Taylor}}, \bibinfo {author}
  {\bibfnamefont {P.}~\bibnamefont {Cappellaro}}, \bibinfo {author}
  {\bibfnamefont {L.}~\bibnamefont {Jiang}}, \bibinfo {author} {\bibfnamefont
  {M.~V.~G.}\ \bibnamefont {Dutt}}, \bibinfo {author} {\bibfnamefont
  {E.}~\bibnamefont {Togan}}, \bibinfo {author} {\bibfnamefont {A.~S.}\
  \bibnamefont {Zibrov}}, \bibinfo {author} {\bibfnamefont {A.}~\bibnamefont
  {Yacoby}}, \bibinfo {author} {\bibfnamefont {R.~L.}\ \bibnamefont
  {Walsworth}}, \ and\ \bibinfo {author} {\bibfnamefont {M.~D.}\ \bibnamefont
  {Lukin}},\ }\href {http://dx.doi.org/10.1038/nature07279} {\bibfield
  {journal} {\bibinfo  {journal} {Nature}\ }\textbf {\bibinfo {volume} {455}},\
  \bibinfo {pages} {644} (\bibinfo {year} {2008})}\BibitemShut {NoStop}%
\bibitem [{\citenamefont {Kucsko}\ \emph {et~al.}(2013)\citenamefont {Kucsko},
  \citenamefont {Maurer}, \citenamefont {Yao}, \citenamefont {Kubo},
  \citenamefont {Noh}, \citenamefont {Lo}, \citenamefont {Park},\ and\
  \citenamefont {Lukin}}]{Kucsko2013kx}%
  \BibitemOpen
  \bibfield  {author} {\bibinfo {author} {\bibfnamefont {G.}~\bibnamefont
  {Kucsko}}, \bibinfo {author} {\bibfnamefont {P.~C.}\ \bibnamefont {Maurer}},
  \bibinfo {author} {\bibfnamefont {N.~Y.}\ \bibnamefont {Yao}}, \bibinfo
  {author} {\bibfnamefont {M.}~\bibnamefont {Kubo}}, \bibinfo {author}
  {\bibfnamefont {H.~J.}\ \bibnamefont {Noh}}, \bibinfo {author} {\bibfnamefont
  {P.~K.}\ \bibnamefont {Lo}}, \bibinfo {author} {\bibfnamefont
  {H.}~\bibnamefont {Park}}, \ and\ \bibinfo {author} {\bibfnamefont {M.~D.}\
  \bibnamefont {Lukin}},\ }\href {http://dx.doi.org/10.1038/nature12373}
  {\bibfield  {journal} {\bibinfo  {journal} {Nature}\ }\textbf {\bibinfo
  {volume} {500}},\ \bibinfo {pages} {54} (\bibinfo {year} {2013})}\BibitemShut
  {NoStop}%
\bibitem [{\citenamefont {Toyli}\ \emph {et~al.}(2013)\citenamefont {Toyli},
  \citenamefont {de~las Casas}, \citenamefont {Christle}, \citenamefont
  {Dobrovitski},\ and\ \citenamefont {Awschalom}}]{Toyli21052013}%
  \BibitemOpen
  \bibfield  {author} {\bibinfo {author} {\bibfnamefont {D.~M.}\ \bibnamefont
  {Toyli}}, \bibinfo {author} {\bibfnamefont {C.~F.}\ \bibnamefont {de~las
  Casas}}, \bibinfo {author} {\bibfnamefont {D.~J.}\ \bibnamefont {Christle}},
  \bibinfo {author} {\bibfnamefont {V.~V.}\ \bibnamefont {Dobrovitski}}, \ and\
  \bibinfo {author} {\bibfnamefont {D.~D.}\ \bibnamefont {Awschalom}},\ }\href
  {\doibase 10.1073/pnas.1306825110} {\bibfield  {journal} {\bibinfo  {journal}
  {Proceedings of the National Academy of Sciences}\ }\textbf {\bibinfo
  {volume} {110}},\ \bibinfo {pages} {8417} (\bibinfo {year}
  {2013})}\BibitemShut {NoStop}%
\bibitem [{\citenamefont {Neumann}\ \emph
  {et~al.}(2010{\natexlab{a}})\citenamefont {Neumann}, \citenamefont {Kolesov},
  \citenamefont {Naydenov}, \citenamefont {Beck}, \citenamefont {Rempp},
  \citenamefont {Steiner}, \citenamefont {Jacques}, \citenamefont
  {Balasubramanian}, \citenamefont {Markham}, \citenamefont {Twitchen},
  \citenamefont {Pezzagna}, \citenamefont {Meijer}, \citenamefont {Twamley},
  \citenamefont {Jelezko},\ and\ \citenamefont {Wrachtrup}}]{Neumann:2010la}%
  \BibitemOpen
  \bibfield  {author} {\bibinfo {author} {\bibfnamefont {P.}~\bibnamefont
  {Neumann}}, \bibinfo {author} {\bibfnamefont {R.}~\bibnamefont {Kolesov}},
  \bibinfo {author} {\bibfnamefont {B.}~\bibnamefont {Naydenov}}, \bibinfo
  {author} {\bibfnamefont {J.}~\bibnamefont {Beck}}, \bibinfo {author}
  {\bibfnamefont {F.}~\bibnamefont {Rempp}}, \bibinfo {author} {\bibfnamefont
  {M.}~\bibnamefont {Steiner}}, \bibinfo {author} {\bibfnamefont
  {V.}~\bibnamefont {Jacques}}, \bibinfo {author} {\bibfnamefont
  {G.}~\bibnamefont {Balasubramanian}}, \bibinfo {author} {\bibfnamefont
  {M.~L.}\ \bibnamefont {Markham}}, \bibinfo {author} {\bibfnamefont {D.~J.}\
  \bibnamefont {Twitchen}}, \bibinfo {author} {\bibfnamefont {S.}~\bibnamefont
  {Pezzagna}}, \bibinfo {author} {\bibfnamefont {J.}~\bibnamefont {Meijer}},
  \bibinfo {author} {\bibfnamefont {J.}~\bibnamefont {Twamley}}, \bibinfo
  {author} {\bibfnamefont {F.}~\bibnamefont {Jelezko}}, \ and\ \bibinfo
  {author} {\bibfnamefont {J.}~\bibnamefont {Wrachtrup}},\ }\href
  {http://dx.doi.org/10.1038/nphys1536} {\bibfield  {journal} {\bibinfo
  {journal} {Nat Phys}\ }\textbf {\bibinfo {volume} {6}},\ \bibinfo {pages}
  {249} (\bibinfo {year} {2010}{\natexlab{a}})}\BibitemShut {NoStop}%
\bibitem [{\citenamefont {Dutt}\ \emph {et~al.}(2007)\citenamefont {Dutt},
  \citenamefont {Childress}, \citenamefont {Jiang}, \citenamefont {Togan},
  \citenamefont {Maze}, \citenamefont {Jelezko}, \citenamefont {Zibrov},
  \citenamefont {Hemmer},\ and\ \citenamefont {Lukin}}]{Dutt01062007}%
  \BibitemOpen
  \bibfield  {author} {\bibinfo {author} {\bibfnamefont {M.~V.~G.}\
  \bibnamefont {Dutt}}, \bibinfo {author} {\bibfnamefont {L.}~\bibnamefont
  {Childress}}, \bibinfo {author} {\bibfnamefont {L.}~\bibnamefont {Jiang}},
  \bibinfo {author} {\bibfnamefont {E.}~\bibnamefont {Togan}}, \bibinfo
  {author} {\bibfnamefont {J.}~\bibnamefont {Maze}}, \bibinfo {author}
  {\bibfnamefont {F.}~\bibnamefont {Jelezko}}, \bibinfo {author} {\bibfnamefont
  {A.~S.}\ \bibnamefont {Zibrov}}, \bibinfo {author} {\bibfnamefont {P.~R.}\
  \bibnamefont {Hemmer}}, \ and\ \bibinfo {author} {\bibfnamefont {M.~D.}\
  \bibnamefont {Lukin}},\ }\href {\doibase 10.1126/science.1139831} {\bibfield
  {journal} {\bibinfo  {journal} {Science}\ }\textbf {\bibinfo {volume}
  {316}},\ \bibinfo {pages} {1312} (\bibinfo {year} {2007})}\BibitemShut
  {NoStop}%
\bibitem [{\citenamefont {van~der Sar}\ \emph {et~al.}(2012)\citenamefont
  {van~der Sar}, \citenamefont {Wang}, \citenamefont {Blok}, \citenamefont
  {Bernien}, \citenamefont {Taminiau}, \citenamefont {Toyli}, \citenamefont
  {Lidar}, \citenamefont {Awschalom}, \citenamefont {Hanson},\ and\
  \citenamefont {Dobrovitski}}]{vanderSar2012}%
  \BibitemOpen
  \bibfield  {author} {\bibinfo {author} {\bibfnamefont {T.}~\bibnamefont
  {van~der Sar}}, \bibinfo {author} {\bibfnamefont {Z.~H.}\ \bibnamefont
  {Wang}}, \bibinfo {author} {\bibfnamefont {M.~S.}\ \bibnamefont {Blok}},
  \bibinfo {author} {\bibfnamefont {H.}~\bibnamefont {Bernien}}, \bibinfo
  {author} {\bibfnamefont {T.~H.}\ \bibnamefont {Taminiau}}, \bibinfo {author}
  {\bibfnamefont {D.~M.}\ \bibnamefont {Toyli}}, \bibinfo {author}
  {\bibfnamefont {D.~A.}\ \bibnamefont {Lidar}}, \bibinfo {author}
  {\bibfnamefont {D.~D.}\ \bibnamefont {Awschalom}}, \bibinfo {author}
  {\bibfnamefont {R.}~\bibnamefont {Hanson}}, \ and\ \bibinfo {author}
  {\bibfnamefont {V.~V.}\ \bibnamefont {Dobrovitski}},\ }\href
  {http://dx.doi.org/10.1038/nature10900} {\bibfield  {journal} {\bibinfo
  {journal} {Nature}\ }\textbf {\bibinfo {volume} {484}},\ \bibinfo {pages}
  {82} (\bibinfo {year} {2012})}\BibitemShut {NoStop}%
\bibitem [{\citenamefont {Jiang}\ \emph {et~al.}(2009)\citenamefont {Jiang},
  \citenamefont {Hodges}, \citenamefont {Maze}, \citenamefont {Maurer},
  \citenamefont {Taylor}, \citenamefont {Cory}, \citenamefont {Hemmer},
  \citenamefont {Walsworth}, \citenamefont {Yacoby}, \citenamefont {Zibrov},\
  and\ \citenamefont {Lukin}}]{Jiang09102009}%
  \BibitemOpen
  \bibfield  {author} {\bibinfo {author} {\bibfnamefont {L.}~\bibnamefont
  {Jiang}}, \bibinfo {author} {\bibfnamefont {J.~S.}\ \bibnamefont {Hodges}},
  \bibinfo {author} {\bibfnamefont {J.~R.}\ \bibnamefont {Maze}}, \bibinfo
  {author} {\bibfnamefont {P.}~\bibnamefont {Maurer}}, \bibinfo {author}
  {\bibfnamefont {J.~M.}\ \bibnamefont {Taylor}}, \bibinfo {author}
  {\bibfnamefont {D.~G.}\ \bibnamefont {Cory}}, \bibinfo {author}
  {\bibfnamefont {P.~R.}\ \bibnamefont {Hemmer}}, \bibinfo {author}
  {\bibfnamefont {R.~L.}\ \bibnamefont {Walsworth}}, \bibinfo {author}
  {\bibfnamefont {A.}~\bibnamefont {Yacoby}}, \bibinfo {author} {\bibfnamefont
  {A.~S.}\ \bibnamefont {Zibrov}}, \ and\ \bibinfo {author} {\bibfnamefont
  {M.~D.}\ \bibnamefont {Lukin}},\ }\href {\doibase 10.1126/science.1176496}
  {\bibfield  {journal} {\bibinfo  {journal} {Science}\ }\textbf {\bibinfo
  {volume} {326}},\ \bibinfo {pages} {267} (\bibinfo {year}
  {2009})}\BibitemShut {NoStop}%
\bibitem [{\citenamefont {Neumann}\ \emph
  {et~al.}(2010{\natexlab{b}})\citenamefont {Neumann}, \citenamefont {Beck},
  \citenamefont {Steiner}, \citenamefont {Rempp}, \citenamefont {Fedder},
  \citenamefont {Hemmer}, \citenamefont {Wrachtrup},\ and\ \citenamefont
  {Jelezko}}]{PhilippNeumann07302010}%
  \BibitemOpen
  \bibfield  {author} {\bibinfo {author} {\bibfnamefont {P.}~\bibnamefont
  {Neumann}}, \bibinfo {author} {\bibfnamefont {J.}~\bibnamefont {Beck}},
  \bibinfo {author} {\bibfnamefont {M.}~\bibnamefont {Steiner}}, \bibinfo
  {author} {\bibfnamefont {F.}~\bibnamefont {Rempp}}, \bibinfo {author}
  {\bibfnamefont {H.}~\bibnamefont {Fedder}}, \bibinfo {author} {\bibfnamefont
  {P.~R.}\ \bibnamefont {Hemmer}}, \bibinfo {author} {\bibfnamefont
  {J.}~\bibnamefont {Wrachtrup}}, \ and\ \bibinfo {author} {\bibfnamefont
  {F.}~\bibnamefont {Jelezko}},\ }\href {\doibase 10.1126/science.1189075}
  {\bibfield  {journal} {\bibinfo  {journal} {Science}\ }\textbf {\bibinfo
  {volume} {329}},\ \bibinfo {pages} {542} (\bibinfo {year}
  {2010}{\natexlab{b}})}\BibitemShut {NoStop}%
\bibitem [{\citenamefont {Robledo}\ \emph
  {et~al.}(2011{\natexlab{a}})\citenamefont {Robledo}, \citenamefont
  {Childress}, \citenamefont {Bernien}, \citenamefont {Hensen}, \citenamefont
  {Alkemade},\ and\ \citenamefont {Hanson}}]{Robledo:2011sf}%
  \BibitemOpen
  \bibfield  {author} {\bibinfo {author} {\bibfnamefont {L.}~\bibnamefont
  {Robledo}}, \bibinfo {author} {\bibfnamefont {L.}~\bibnamefont {Childress}},
  \bibinfo {author} {\bibfnamefont {H.}~\bibnamefont {Bernien}}, \bibinfo
  {author} {\bibfnamefont {B.}~\bibnamefont {Hensen}}, \bibinfo {author}
  {\bibfnamefont {P.~F.~A.}\ \bibnamefont {Alkemade}}, \ and\ \bibinfo {author}
  {\bibfnamefont {R.}~\bibnamefont {Hanson}},\ }\href
  {http://dx.doi.org/10.1038/nature10401} {\bibfield  {journal} {\bibinfo
  {journal} {Nature}\ }\textbf {\bibinfo {volume} {477}},\ \bibinfo {pages}
  {574} (\bibinfo {year} {2011}{\natexlab{a}})}\BibitemShut {NoStop}%
\bibitem [{\citenamefont {Han}\ \emph {et~al.}(2010)\citenamefont {Han},
  \citenamefont {Kim}, \citenamefont {Eggeling},\ and\ \citenamefont
  {Hell}}]{doi:10.1021/nl102156m}%
  \BibitemOpen
  \bibfield  {author} {\bibinfo {author} {\bibfnamefont {K.~Y.}\ \bibnamefont
  {Han}}, \bibinfo {author} {\bibfnamefont {S.~K.}\ \bibnamefont {Kim}},
  \bibinfo {author} {\bibfnamefont {C.}~\bibnamefont {Eggeling}}, \ and\
  \bibinfo {author} {\bibfnamefont {S.~W.}\ \bibnamefont {Hell}},\ }\href
  {\doibase 10.1021/nl102156m} {\bibfield  {journal} {\bibinfo  {journal} {Nano
  Letters}\ }\textbf {\bibinfo {volume} {10}},\ \bibinfo {pages} {3199}
  (\bibinfo {year} {2010})}\BibitemShut {NoStop}%
\bibitem [{\citenamefont {Han}\ \emph {et~al.}(2012)\citenamefont {Han},
  \citenamefont {Wildanger}, \citenamefont {Rittweger}, \citenamefont {Meijer},
  \citenamefont {Pezzagna}, \citenamefont {Hell},\ and\ \citenamefont
  {Eggeling}}]{1367-2630-14-12-123002}%
  \BibitemOpen
  \bibfield  {author} {\bibinfo {author} {\bibfnamefont {K.~Y.}\ \bibnamefont
  {Han}}, \bibinfo {author} {\bibfnamefont {D.}~\bibnamefont {Wildanger}},
  \bibinfo {author} {\bibfnamefont {E.}~\bibnamefont {Rittweger}}, \bibinfo
  {author} {\bibfnamefont {J.}~\bibnamefont {Meijer}}, \bibinfo {author}
  {\bibfnamefont {S.}~\bibnamefont {Pezzagna}}, \bibinfo {author}
  {\bibfnamefont {S.~W.}\ \bibnamefont {Hell}}, \ and\ \bibinfo {author}
  {\bibfnamefont {C.}~\bibnamefont {Eggeling}},\ }\href
  {http://stacks.iop.org/1367-2630/14/i=12/a=123002} {\bibfield  {journal}
  {\bibinfo  {journal} {New Journal of Physics}\ }\textbf {\bibinfo {volume}
  {14}},\ \bibinfo {pages} {123002} (\bibinfo {year} {2012})}\BibitemShut
  {NoStop}%
\bibitem [{\citenamefont {Waldherr}\ \emph {et~al.}(2011)\citenamefont
  {Waldherr}, \citenamefont {Beck}, \citenamefont {Steiner}, \citenamefont
  {Neumann}, \citenamefont {Gali}, \citenamefont {Frauenheim}, \citenamefont
  {Jelezko},\ and\ \citenamefont {Wrachtrup}}]{PhysRevLett.106.157601}%
  \BibitemOpen
  \bibfield  {author} {\bibinfo {author} {\bibfnamefont {G.}~\bibnamefont
  {Waldherr}}, \bibinfo {author} {\bibfnamefont {J.}~\bibnamefont {Beck}},
  \bibinfo {author} {\bibfnamefont {M.}~\bibnamefont {Steiner}}, \bibinfo
  {author} {\bibfnamefont {P.}~\bibnamefont {Neumann}}, \bibinfo {author}
  {\bibfnamefont {A.}~\bibnamefont {Gali}}, \bibinfo {author} {\bibfnamefont
  {T.}~\bibnamefont {Frauenheim}}, \bibinfo {author} {\bibfnamefont
  {F.}~\bibnamefont {Jelezko}}, \ and\ \bibinfo {author} {\bibfnamefont
  {J.}~\bibnamefont {Wrachtrup}},\ }\href {\doibase
  10.1103/PhysRevLett.106.157601} {\bibfield  {journal} {\bibinfo  {journal}
  {Phys. Rev. Lett.}\ }\textbf {\bibinfo {volume} {106}},\ \bibinfo {pages}
  {157601} (\bibinfo {year} {2011})}\BibitemShut {NoStop}%
\bibitem [{\citenamefont {Beha}\ \emph {et~al.}(2012)\citenamefont {Beha},
  \citenamefont {Batalov}, \citenamefont {Manson}, \citenamefont
  {Bratschitsch},\ and\ \citenamefont
  {Leitenstorfer}}]{PhysRevLett.109.097404}%
  \BibitemOpen
  \bibfield  {author} {\bibinfo {author} {\bibfnamefont {K.}~\bibnamefont
  {Beha}}, \bibinfo {author} {\bibfnamefont {A.}~\bibnamefont {Batalov}},
  \bibinfo {author} {\bibfnamefont {N.~B.}\ \bibnamefont {Manson}}, \bibinfo
  {author} {\bibfnamefont {R.}~\bibnamefont {Bratschitsch}}, \ and\ \bibinfo
  {author} {\bibfnamefont {A.}~\bibnamefont {Leitenstorfer}},\ }\href {\doibase
  10.1103/PhysRevLett.109.097404} {\bibfield  {journal} {\bibinfo  {journal}
  {Phys. Rev. Lett.}\ }\textbf {\bibinfo {volume} {109}},\ \bibinfo {pages}
  {097404} (\bibinfo {year} {2012})}\BibitemShut {NoStop}%
\bibitem [{\citenamefont {Aslam}\ \emph {et~al.}(2013)\citenamefont {Aslam},
  \citenamefont {Waldherr}, \citenamefont {Neumann}, \citenamefont {Jelezko},\
  and\ \citenamefont {Wrachtrup}}]{1367-2630-15-1-013064}%
  \BibitemOpen
  \bibfield  {author} {\bibinfo {author} {\bibfnamefont {N.}~\bibnamefont
  {Aslam}}, \bibinfo {author} {\bibfnamefont {G.}~\bibnamefont {Waldherr}},
  \bibinfo {author} {\bibfnamefont {P.}~\bibnamefont {Neumann}}, \bibinfo
  {author} {\bibfnamefont {F.}~\bibnamefont {Jelezko}}, \ and\ \bibinfo
  {author} {\bibfnamefont {J.}~\bibnamefont {Wrachtrup}},\ }\href
  {http://stacks.iop.org/1367-2630/15/i=1/a=013064} {\bibfield  {journal}
  {\bibinfo  {journal} {New Journal of Physics}\ }\textbf {\bibinfo {volume}
  {15}},\ \bibinfo {pages} {013064} (\bibinfo {year} {2013})}\BibitemShut
  {NoStop}%
\bibitem [{\citenamefont {Manson}\ and\ \citenamefont
  {Harrison}(2005)}]{Manson20051705}%
  \BibitemOpen
  \bibfield  {author} {\bibinfo {author} {\bibfnamefont {N.}~\bibnamefont
  {Manson}}\ and\ \bibinfo {author} {\bibfnamefont {J.}~\bibnamefont
  {Harrison}},\ }\href {\doibase
  http://dx.doi.org/10.1016/j.diamond.2005.06.027} {\bibfield  {journal}
  {\bibinfo  {journal} {Diamond and Related Materials}\ }\textbf {\bibinfo
  {volume} {14}},\ \bibinfo {pages} {1705 } (\bibinfo {year}
  {2005})}\BibitemShut {NoStop}%
\bibitem [{\citenamefont {Gruber}\ \emph {et~al.}(1997)\citenamefont {Gruber},
  \citenamefont {Dr{\"a}benstedt}, \citenamefont {Tietz}, \citenamefont
  {Fleury}, \citenamefont {Wrachtrup},\ and\ \citenamefont
  {Borczyskowski}}]{Gruber27061997}%
  \BibitemOpen
  \bibfield  {author} {\bibinfo {author} {\bibfnamefont {A.}~\bibnamefont
  {Gruber}}, \bibinfo {author} {\bibfnamefont {A.}~\bibnamefont
  {Dr{\"a}benstedt}}, \bibinfo {author} {\bibfnamefont {C.}~\bibnamefont
  {Tietz}}, \bibinfo {author} {\bibfnamefont {L.}~\bibnamefont {Fleury}},
  \bibinfo {author} {\bibfnamefont {J.}~\bibnamefont {Wrachtrup}}, \ and\
  \bibinfo {author} {\bibfnamefont {C.~v.}\ \bibnamefont {Borczyskowski}},\
  }\href {\doibase 10.1126/science.276.5321.2012} {\bibfield  {journal}
  {\bibinfo  {journal} {Science}\ }\textbf {\bibinfo {volume} {276}},\ \bibinfo
  {pages} {2012} (\bibinfo {year} {1997})}\BibitemShut {NoStop}%
\bibitem [{\citenamefont {Burek}\ \emph {et~al.}(2012)\citenamefont {Burek},
  \citenamefont {de~Leon}, \citenamefont {Shields}, \citenamefont {Hausmann},
  \citenamefont {Chu}, \citenamefont {Quan}, \citenamefont {Zibrov},
  \citenamefont {Park}, \citenamefont {Lukin},\ and\ \citenamefont {Lon{\v
  c}ar}}]{doi:10.1021/nl302541e}%
  \BibitemOpen
  \bibfield  {author} {\bibinfo {author} {\bibfnamefont {M.~J.}\ \bibnamefont
  {Burek}}, \bibinfo {author} {\bibfnamefont {N.~P.}\ \bibnamefont {de~Leon}},
  \bibinfo {author} {\bibfnamefont {B.~J.}\ \bibnamefont {Shields}}, \bibinfo
  {author} {\bibfnamefont {B.~J.~M.}\ \bibnamefont {Hausmann}}, \bibinfo
  {author} {\bibfnamefont {Y.}~\bibnamefont {Chu}}, \bibinfo {author}
  {\bibfnamefont {Q.}~\bibnamefont {Quan}}, \bibinfo {author} {\bibfnamefont
  {A.~S.}\ \bibnamefont {Zibrov}}, \bibinfo {author} {\bibfnamefont
  {H.}~\bibnamefont {Park}}, \bibinfo {author} {\bibfnamefont {M.~D.}\
  \bibnamefont {Lukin}}, \ and\ \bibinfo {author} {\bibfnamefont
  {M.}~\bibnamefont {Lon{\v c}ar}},\ }\href {\doibase 10.1021/nl302541e}
  {\bibfield  {journal} {\bibinfo  {journal} {Nano Letters}\ }\textbf {\bibinfo
  {volume} {12}},\ \bibinfo {pages} {6084} (\bibinfo {year}
  {2012})}\BibitemShut {NoStop}%
\bibitem [{\citenamefont {Lee}\ \emph {et~al.}(2011)\citenamefont {Lee},
  \citenamefont {Chen}, \citenamefont {Eghlidi}, \citenamefont {Kukura},
  \citenamefont {Lettow}, \citenamefont {Renn}, \citenamefont {Sandoghdar},\
  and\ \citenamefont {Gotzinger}}]{G2011uq}%
  \BibitemOpen
  \bibfield  {author} {\bibinfo {author} {\bibfnamefont {K.~G.}\ \bibnamefont
  {Lee}}, \bibinfo {author} {\bibfnamefont {X.~W.}\ \bibnamefont {Chen}},
  \bibinfo {author} {\bibfnamefont {H.}~\bibnamefont {Eghlidi}}, \bibinfo
  {author} {\bibfnamefont {P.}~\bibnamefont {Kukura}}, \bibinfo {author}
  {\bibfnamefont {R.}~\bibnamefont {Lettow}}, \bibinfo {author} {\bibfnamefont
  {A.}~\bibnamefont {Renn}}, \bibinfo {author} {\bibfnamefont {V.}~\bibnamefont
  {Sandoghdar}}, \ and\ \bibinfo {author} {\bibfnamefont {S.}~\bibnamefont
  {Gotzinger}},\ }\href {http://dx.doi.org/10.1038/nphoton.2010.312} {\bibfield
   {journal} {\bibinfo  {journal} {Nat Photon}\ }\textbf {\bibinfo {volume}
  {5}},\ \bibinfo {pages} {166} (\bibinfo {year} {2011})}\BibitemShut {NoStop}%
\bibitem [{\citenamefont {Riedel}\ \emph {et~al.}(2014)\citenamefont {Riedel},
  \citenamefont {Rohner}, \citenamefont {Ganzhorn}, \citenamefont {Kaldewey},
  \citenamefont {Appel}, \citenamefont {Neu}, \citenamefont {Warburton},\ and\
  \citenamefont {Maletinsky}}]{Maletinsky2014}%
  \BibitemOpen
  \bibfield  {author} {\bibinfo {author} {\bibfnamefont {D.}~\bibnamefont
  {Riedel}}, \bibinfo {author} {\bibfnamefont {D.}~\bibnamefont {Rohner}},
  \bibinfo {author} {\bibfnamefont {M.}~\bibnamefont {Ganzhorn}}, \bibinfo
  {author} {\bibfnamefont {T.}~\bibnamefont {Kaldewey}}, \bibinfo {author}
  {\bibfnamefont {P.}~\bibnamefont {Appel}}, \bibinfo {author} {\bibfnamefont
  {E.}~\bibnamefont {Neu}}, \bibinfo {author} {\bibfnamefont {R.~J.}\
  \bibnamefont {Warburton}}, \ and\ \bibinfo {author} {\bibfnamefont
  {P.}~\bibnamefont {Maletinsky}},\ }\href@noop {} {\bibfield  {journal}
  {\bibinfo  {journal} {arXiv:1408.4117 [cond-mat.mes-hall]}\ } (\bibinfo
  {year} {2014})}\BibitemShut {NoStop}%
\bibitem [{sup()}]{sup}%
  \BibitemOpen
  \href@noop {} {\bibinfo  {journal} {See Supplemental Information}\
  }\BibitemShut {NoStop}%
\bibitem [{\citenamefont {Taylor}\ \emph {et~al.}(2008)\citenamefont {Taylor},
  \citenamefont {Cappellaro}, \citenamefont {Childress}, \citenamefont {Jiang},
  \citenamefont {Budker}, \citenamefont {Hemmer}, \citenamefont {Yacoby},
  \citenamefont {Walsworth},\ and\ \citenamefont {Lukin}}]{Taylor:2008nx}%
  \BibitemOpen
\bibfield  {journal} {  }\bibfield  {author} {\bibinfo {author} {\bibfnamefont
  {J.~M.}\ \bibnamefont {Taylor}}, \bibinfo {author} {\bibfnamefont
  {P.}~\bibnamefont {Cappellaro}}, \bibinfo {author} {\bibfnamefont
  {L.}~\bibnamefont {Childress}}, \bibinfo {author} {\bibfnamefont
  {L.}~\bibnamefont {Jiang}}, \bibinfo {author} {\bibfnamefont
  {D.}~\bibnamefont {Budker}}, \bibinfo {author} {\bibfnamefont {P.~R.}\
  \bibnamefont {Hemmer}}, \bibinfo {author} {\bibfnamefont {A.}~\bibnamefont
  {Yacoby}}, \bibinfo {author} {\bibfnamefont {R.}~\bibnamefont {Walsworth}}, \
  and\ \bibinfo {author} {\bibfnamefont {M.~D.}\ \bibnamefont {Lukin}},\ }\href
  {http://dx.doi.org/10.1038/nphys1075} {\bibfield  {journal} {\bibinfo
  {journal} {Nat Phys}\ }\textbf {\bibinfo {volume} {4}},\ \bibinfo {pages}
  {810} (\bibinfo {year} {2008})}\BibitemShut {NoStop}%
\bibitem [{\citenamefont {Hahn}(1950)}]{PhysRev.80.580}%
  \BibitemOpen
  \bibfield  {author} {\bibinfo {author} {\bibfnamefont {E.~L.}\ \bibnamefont
  {Hahn}},\ }\href {\doibase 10.1103/PhysRev.80.580} {\bibfield  {journal}
  {\bibinfo  {journal} {Phys. Rev.}\ }\textbf {\bibinfo {volume} {80}},\
  \bibinfo {pages} {580} (\bibinfo {year} {1950})}\BibitemShut {NoStop}%
\bibitem [{\citenamefont {Balasubramanian}\ \emph {et~al.}(2009)\citenamefont
  {Balasubramanian}, \citenamefont {Neumann}, \citenamefont {Twitchen},
  \citenamefont {Markham}, \citenamefont {Kolesov}, \citenamefont {Mizuochi},
  \citenamefont {Isoya}, \citenamefont {Achard}, \citenamefont {Beck},
  \citenamefont {Tissler}, \citenamefont {Jacques}, \citenamefont {Hemmer},
  \citenamefont {Jelezko},\ and\ \citenamefont
  {Wrachtrup}}]{Balasubramanian:2009bf}%
  \BibitemOpen
  \bibfield  {author} {\bibinfo {author} {\bibfnamefont {G.}~\bibnamefont
  {Balasubramanian}}, \bibinfo {author} {\bibfnamefont {P.}~\bibnamefont
  {Neumann}}, \bibinfo {author} {\bibfnamefont {D.}~\bibnamefont {Twitchen}},
  \bibinfo {author} {\bibfnamefont {M.}~\bibnamefont {Markham}}, \bibinfo
  {author} {\bibfnamefont {R.}~\bibnamefont {Kolesov}}, \bibinfo {author}
  {\bibfnamefont {N.}~\bibnamefont {Mizuochi}}, \bibinfo {author}
  {\bibfnamefont {J.}~\bibnamefont {Isoya}}, \bibinfo {author} {\bibfnamefont
  {J.}~\bibnamefont {Achard}}, \bibinfo {author} {\bibfnamefont
  {J.}~\bibnamefont {Beck}}, \bibinfo {author} {\bibfnamefont {J.}~\bibnamefont
  {Tissler}}, \bibinfo {author} {\bibfnamefont {V.}~\bibnamefont {Jacques}},
  \bibinfo {author} {\bibfnamefont {P.~R.}\ \bibnamefont {Hemmer}}, \bibinfo
  {author} {\bibfnamefont {F.}~\bibnamefont {Jelezko}}, \ and\ \bibinfo
  {author} {\bibfnamefont {J.}~\bibnamefont {Wrachtrup}},\ }\href
  {http://dx.doi.org/10.1038/nmat2420} {\bibfield  {journal} {\bibinfo
  {journal} {Nat Mater}\ }\textbf {\bibinfo {volume} {8}},\ \bibinfo {pages}
  {383} (\bibinfo {year} {2009})}\BibitemShut {NoStop}%
\bibitem [{\citenamefont {Childress}\ \emph {et~al.}(2006)\citenamefont
  {Childress}, \citenamefont {Gurudev~Dutt}, \citenamefont {Taylor},
  \citenamefont {Zibrov}, \citenamefont {Jelezko}, \citenamefont {Wrachtrup},
  \citenamefont {Hemmer},\ and\ \citenamefont
  {Lukin}}]{2006.Science.Childress.coherent_nuclear_spin_dynamics}%
  \BibitemOpen
  \bibfield  {author} {\bibinfo {author} {\bibfnamefont {L.}~\bibnamefont
  {Childress}}, \bibinfo {author} {\bibfnamefont {M.~V.}\ \bibnamefont
  {Gurudev~Dutt}}, \bibinfo {author} {\bibfnamefont {J.~M.}\ \bibnamefont
  {Taylor}}, \bibinfo {author} {\bibfnamefont {A.~S.}\ \bibnamefont {Zibrov}},
  \bibinfo {author} {\bibfnamefont {F.}~\bibnamefont {Jelezko}}, \bibinfo
  {author} {\bibfnamefont {J.}~\bibnamefont {Wrachtrup}}, \bibinfo {author}
  {\bibfnamefont {P.~R.}\ \bibnamefont {Hemmer}}, \ and\ \bibinfo {author}
  {\bibfnamefont {M.~D.}\ \bibnamefont {Lukin}},\ }\href {\doibase
  10.1126/science.1131871} {\bibfield  {journal} {\bibinfo  {journal}
  {Science}\ }\textbf {\bibinfo {volume} {314}},\ \bibinfo {pages} {281}
  (\bibinfo {year} {2006})}\BibitemShut {NoStop}%
\bibitem [{\citenamefont {Robledo}\ \emph
  {et~al.}(2011{\natexlab{b}})\citenamefont {Robledo}, \citenamefont {Bernien},
  \citenamefont {van~der Sar},\ and\ \citenamefont
  {Hanson}}]{1367-2630-13-2-025013}%
  \BibitemOpen
  \bibfield  {author} {\bibinfo {author} {\bibfnamefont {L.}~\bibnamefont
  {Robledo}}, \bibinfo {author} {\bibfnamefont {H.}~\bibnamefont {Bernien}},
  \bibinfo {author} {\bibfnamefont {T.}~\bibnamefont {van~der Sar}}, \ and\
  \bibinfo {author} {\bibfnamefont {R.}~\bibnamefont {Hanson}},\ }\href
  {http://stacks.iop.org/1367-2630/13/i=2/a=025013} {\bibfield  {journal}
  {\bibinfo  {journal} {New Journal of Physics}\ }\textbf {\bibinfo {volume}
  {13}},\ \bibinfo {pages} {025013} (\bibinfo {year}
  {2011}{\natexlab{b}})}\BibitemShut {NoStop}%
\end{thebibliography}%


\pagebreak
\begin{widetext}
\begin{center}
\textbf{\large Supplemental materials for efficient readout of a single spin state in diamond via spin-to-charge conversion}
\end{center}
\end{widetext}

\setcounter{equation}{0}
\setcounter{figure}{0}
\setcounter{table}{0}
\setcounter{page}{1}
\makeatletter
\renewcommand{\theequation}{S\arabic{equation}}
\renewcommand{\thefigure}{S\arabic{figure}}
\renewcommand{\bibnumfmt}[1]{[S#1]}
\renewcommand{\citenumfont}[1]{S#1}

\section{Device Fabrication}

We employ an angled RIE fabrication technique\cite{doi:10.1021/nl302541e} to carve \SI{300}{\nano\meter}-wide, triangular cross-section nanobeams from a bulk diamond sample.  To do so, we begin with a polished diamond sample (Element6, type IIa, 1ppm N concentration) and remove $\sim$ \SI{600}{\nano\meter} of material from the top surface in a top-down oxygen RIE step.  Next, we spin a PMMA layer onto the diamond and pattern the beam mask shape via e-beam lithography.  After developing the PMMA, a \SI{294}{\nano\meter}-thick layer of Al$_2$O$_3$ is sputtered and the PMMA is stripped, transferring the etch mask pattern into the Al$_2$O$_3$ layer.  Next, we perform a top-down etch for \SI{3}{\minute} in O$_2$ plasma to create vertical clearance for the angled etch.  Following the top-down etch, the angled etch is performed in a Faraday cage with sloped mesh walls for a total of \SI{7}{\minute} in O$_2$ + Cl$_2$ plasma, with the etch broken into 12 cycles of \SI{35}{\second} each.  The diamond is then cleaned in a boiling solution of 1:1:1 perchloric, nitric, and sulfuric acids, and annealed in a 3-stage ramp consisting of \SI{3}{\hour} ramp from room temperature to \SI{400}{\celsius}, annealing at \SI{400}{\celsius} for \SI{4}{\hour}, \SI{3}{\hour} ramp from \SI{400}{\celsius} to \SI{800}{\celsius}, annealing at \SI{800}{\celsius} for \SI{8}{\hour}, \SI{12}{\hour} ramp from \SI{800}{\celsius} to \SI{1200}{\celsius}, anneal at \SI{1200}{\celsius} for \SI{2}{\hour}, ramp down to room temperature.  Following the anneal the diamond is again acid cleaned and baked at \SI{465}{\celsius} in oxygen environment.

\section{NV Spin Coherence}

\begin{figure}
\centering
\includegraphics{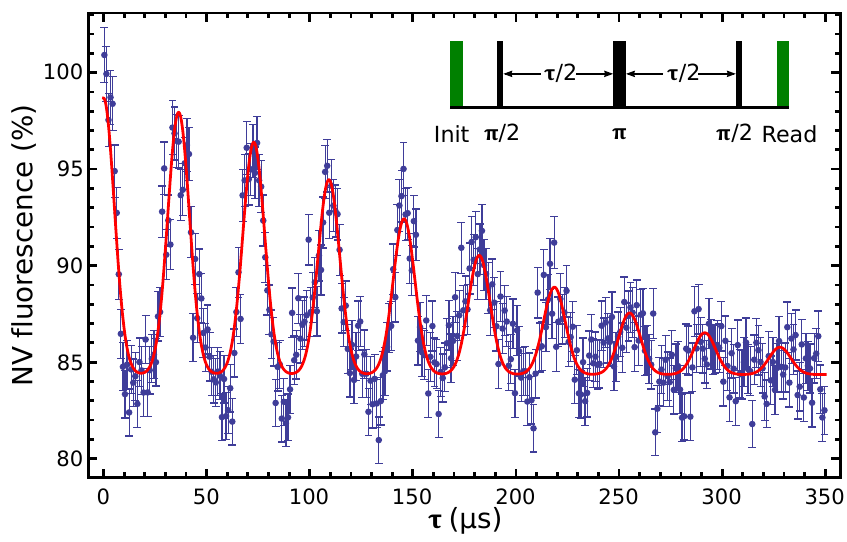}
\caption{Spin echo for NV in nanobeam.  The blue points are fluorescence measured after a Hahn echo, normalized to the fluorescence level with no echo pulses.  The red curve is a fit to the data (see text).}
\label{fig:SCC_SpinEcho}
\end{figure}

We observe similar spin coherence properties in the nanobeams as in bulk, natural $^{13}$C abundance diamond\cite{2006.Science.Childress.coherent_nuclear_spin_dynamics}.  A Hahn echo measurement is shown in Fig.~\ref{fig:SCC_SpinEcho} for a similarly prepared beam as that used for the SCC measurements.  The data is fitted by the function\cite{2006.Science.Childress.coherent_nuclear_spin_dynamics}:
\begin{equation}
	F(\tau) = A + Be^{-(\tau/T_2)^n}\sum_{j=0}^{9} e^{-((\tau-jT_{rev})/T_{dec})^2},
\end{equation} 
with $A=0.844\pm0.001$, $B=0.143\pm0.005$, $n=1.72\pm0.14$, $T_{rev}=36.48\pm0.04\mu$s, $T_{dec}=7.47\pm0.22\mu$s, and $T_2=201\pm7\mu$s.

\section{Model for photon statistics}

\begin{figure}
\centering
\includegraphics{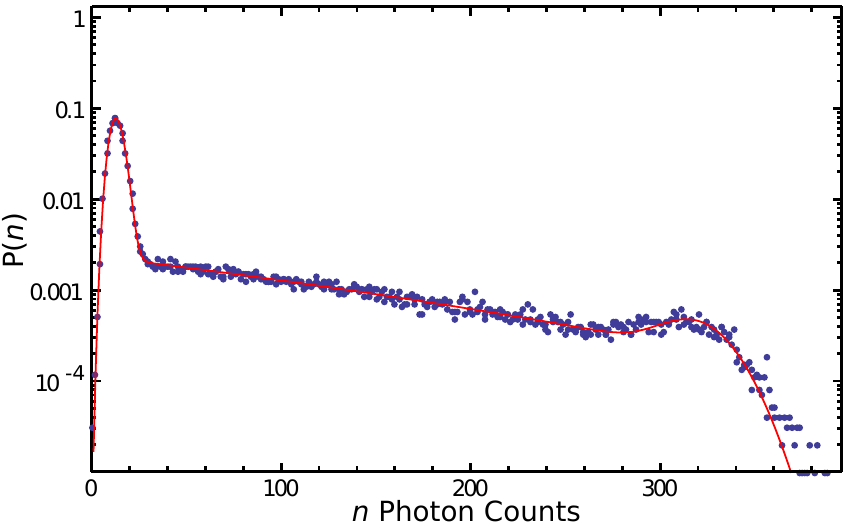}
\caption{Example of model for photon statistics.  Data was taken under \SI{875}{\nano\watt} illumination with cw 594-nm light, integration time \SI{8}{\milli\second}.  100,000 measurements.}
\label{fig:SCC_fit}
\end{figure}

To characterize the charge state quantitatively, we assumed that the dynamics can be fully described by 4 rates: the ionization rates from \nvm to \nvz and vice versa ($g_1$ and $g_0$, respectively), and the photon count rates when in \nvm and \nvz ($\gamma_1$ and $\gamma_0$, respectively).  From $\gamma_{1,0}$, we can calculate the photon number distribution that results from a particular sequence of ionization events.  For example, suppose the NV begins in \nvm, jumps to \nvz after time $\tau_1$, then jumps back to \nvm after an additional time $t_1$ and remains in \nvm for the rest of the counting window.  The photon number distribution for that ionization sequence would be a Poisson distribution with mean value $\gamma_1(t_R-t_1)+\gamma_0t_1$.  The total photon number distribution for a particular initial charge state and total counting time $t_R$ is then a sum over the photon number distributions for all possible ionization sequences, weighted by the probability for each sequence to occur.  In the case that the initial state is \nvm, we have:
\begin{widetext}
\begin{eqnarray}
	p(n|\mathrm{NV}^-,\mathrm{odd}) &=& \int_{0}^{t_R} d\tau e^{(g_0-g_1)\tau-g_0t_R} \sum_{i=1}^{\infty}g_1^ig_0^{i-1} \prod_{j=1}^{i-1}\int_{0}^{\tau -\sum_{k=1}^{(j-1)}\tau_{k}}ds_j \times \nonumber \\
	&& \prod_{j=1}^{i-1}\int_{0}^{(t_R-\tau )-\sum_{k=1}^{(j-1)}t_{k}}dt_j \mathrm{PoissPDF}(\gamma_1\tau+\gamma_0(t_R-\tau),n) \\
	p(n|\mathrm{NV}^-,\mathrm{even}) &=& \int_{0}^{t_R} d\tau e^{(g_0-g_1)\tau-g_0t_R} \sum_{i=1}^{\infty}(g_1g_0)^i \prod_{j=1}^{i}\int_{0}^{\tau-\sum_{k=1}^{(j-1)}\tau_{k}}d\tau_j \times \nonumber \\ 
	&& \prod_{j=1}^{i-1}\int_{0}^{(t_R-\tau)-\sum_{k=1}^{(j-1)}t_{k}}dt_j \mathrm{PoissPDF}(\gamma_1\tau+\gamma_0(t_R-\tau),n) \nonumber \\
	&& + e^{-g_1 t_R}\mathrm{PoissPDF}(\gamma_1 t_R,n)
\end{eqnarray}
\end{widetext}
where $\tau$ is the total time spent in \nvm and must therefore be integrated over $[0,t_R]$, $\mathrm{PoissPDF}(x,n)$ is the probability distribution function for an outcome of $n$ for a Poisson random variable with mean value $x$, and we have broken the result into those cases where there are an odd total number of ionization events and an even number.  The last term in the expression for $p(n|\mathrm{NV}^-,\mathrm{even})$ is the zero ionization event case.  For the case of \nvz as the initial state, simply exchange $1\leftrightarrow0$.  The integral products can be evaluated as the volume of a pyramid in $i$ dimensions, and consequently the sum over ionization events reduces to an expression in terms of Bessel functions:
\begin{widetext}
\begin{eqnarray}
	p(n|\mathrm{NV}^-,\mathrm{odd}) &=& \int_{0}^{t_R} d\tau g_1e^{(g_0-g_1)\tau-g_0t_R} \mathrm{BesselI}(0,2\sqrt{g_1g_0\tau(t_R-\tau)})\mathrm{PoissPDF}(\gamma_1\tau+\gamma_0(t_R-\tau),n) \\
	p(n|\mathrm{NV}^-,\mathrm{even}) &=& \int_{0}^{t_R} d\tau \sqrt{\frac{g_1g_0\tau}{t_R-\tau}}e^{(g_0-g_1)\tau-g_0t_R} \mathrm{BesselI}(1,2\sqrt{g_1g_0\tau(t_R-\tau)})\mathrm{PoissPDF}(\gamma_1\tau+\gamma_0(t_R-\tau),n) \nonumber \\
	&& + e^{-g_1 t_R}\mathrm{PoissPDF}(\gamma_1 t_R,n),
\end{eqnarray}
\end{widetext}
where $\mathrm{BesselI}(m,x)$ is a modified Bessel function of the first kind.  To evaluate these photon number distributions, we performed the integral numerically in \textit{Mathematica}.  This model accurately captures the behavior of the system under the cw, low power illumination conditions used for charge readout, as shown in Fig.~\ref{fig:SCC_fit}

\section{Measuring ionization and photon count rates}

The above model can be used to find the ideal power and time settings that maximize $\mathcal{F}_C$, once the ionization and photon count rates are known.  To measure these, we integrated the counts over a time window $t_R\sim1/g_1(P)$ for a range of cw 594-nm powers, $P$, and used the model to fit the photon number distribution from 100,000 measurements.  The choice of $t_R$ is made to ensure that sufficient ionization events occur to get an accurate fit of the ionization rates.  Since the measurement was steady state, it is sufficient to select $t_R$ to be long enough to measure $g_1$, since $g_{0,1}$ are related by the steady state population balance: $g_0/g_1 = p(NV^-)/p(NV^0)$.
\begin{figure}
\centering
\includegraphics{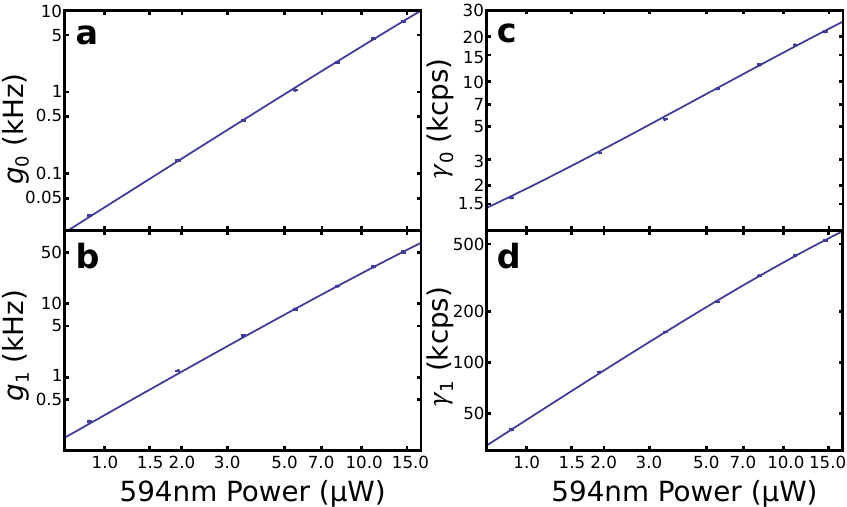}
\caption{Ionization and photon count rates under cw 594-nm illumination.  (a) $g_0$, fitted with a model of the form $aP^2/(1+P/P_{sat})$ with $P_{sat}=\SI{134}{\micro\watt}$, $a=39$ cps/$\mathrm{\mu}$W$^2$.  (b) $g_1$, fitted with a model of the form $aP^2/(1+P/P_{sat})$ with $P_{sat}=\SI{53.2}{\micro\watt}$, $a=310$ cps/$\mathrm{\mu}$W$^2$. (c) $\gamma_0$, fitted with a model of the form $a*P/(1+P/P_{sat})+dc$, with detector dark count rate $dc=0.268$ kcps measured independently, $a=1.65$ kcps/$\mathrm{\mu}$W, $P_{sat}=\SI{134}{\micro\watt}$. (d) $\gamma_1$, fitted with a model of the form $a*P/(1+P/P_{sat})+dc$, with $a=46.2$ kcps/$\mathrm{\mu}$W, $P_{sat}=\SI{53}{\micro\watt}$.  100,000 measurements were taken at all power levels.  Saturation power levels were determined from fits to the $\gamma_{0,1}$ data and the fitted values were then used for the fits of $g_{0,1}$.}
\label{fig:SCC_rates}
\end{figure}

The rates measured via the above fitting procedure for cw 594-nm powers from \SI{875}{\nano\watt} to \SI{14.5}{\micro\watt} are shown in Fig.~\ref{fig:SCC_rates} (blue points).  The count rates are fitted by an expression of the form $a*P/(1+P/P_{sat})+dc$ where $P_{sat}$ is the saturation power and $dc$ is the detector dark count rate, measured to be 0.268 kcps.  The ionization rates are fitted by an expression of the form $aP^2/(1+P/P_{sat})$, where $P_{sat}$ is taken from the corresponding photon count rate fit.

\begin{figure}
\centering
\includegraphics{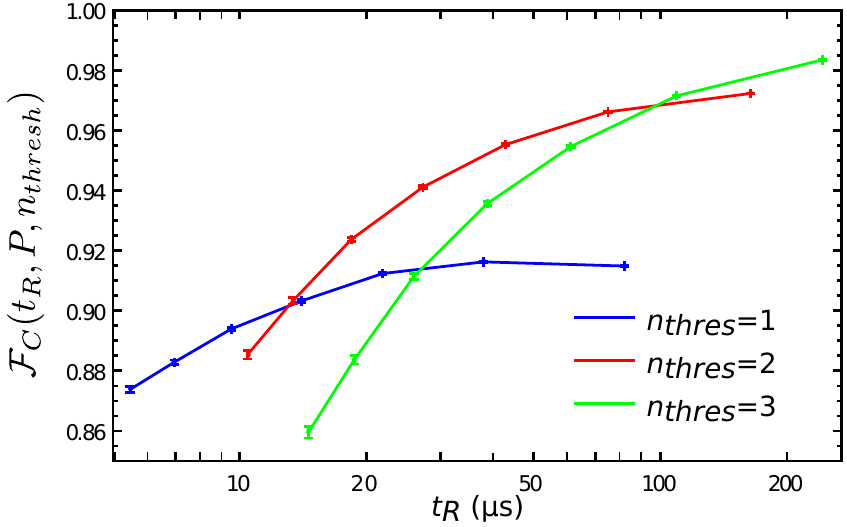}
\caption{Calculated optimal $\mathcal{F}_C$ for a range of powers at photon thresholds $n_{thresh}=[1,2,3]$.  For each power and threshold combination, the readout time is optimized to maximize $\mathcal{F}_C$.}
\label{fig:SCC_Optimal_tR}
\end{figure}

Having measured the rates, we proceed to determine the optimal readout times and corresponding $\mathcal{F}_C(P,t_R)$ for the set of 594-nm powers used.  To do so, we use a simple thresholding algorithm ($n\ge n_{thresh} \rightarrow NV^-$, $n<n_{thresh} \rightarrow NV^0$) under the assumption of a 50/50 charge state population balance.  Then, using the photon distribution from the above model, we calculate the probability of correctly determining the charge state, and maximize that outcome with respect to $t_R$ for each power, using a photon threshold of $n_{thresh}=[1,2,3]$.  The resulting 3 data sets (one for each photon threshold) are shown in Fig.~\ref{fig:SCC_Optimal_tR}.  We use the optimal threshold at each value of $t_R$ for the plot in Fig.~2b.

\section{Readout noise and magnetometer sensitivity}

We consider the following scheme for sensing AC magnetic fields with an NV:
\begin{enumerate}
\item  initialize the NV into $|m_s=0\rangle$, in time $t_I$,
\item  carry out a Hahn echo pulse sequence occupying time $\tau$,
\item read out the NV spin, in time $t_R$.
\end{enumerate}

\noindent At the end of the echo sequence, the state of the system is 
\begin{equation}
\label{eq:SpinEchoState}
	|\psi(\tau)\rangle = \cos\left(\frac{g\mu_B B \tau}{\pi\hbar}\right)|\mathrm{m_s}=0\rangle - i\sin\left(\frac{g\mu_B B \tau}{\pi\hbar}\right)|\mathrm{m_s}=1\rangle,
\end{equation}
where $g$ is the electron Land\'e $g$ factor and $B$ is the magnitude of the magnetic field.  The measurement procedure projects onto one or another of $|\mathrm{m_s}=0\rangle,|\mathrm{m_s}=1\rangle$, with probabilities:
\begin{eqnarray}
	p_0 &=& \cos^2\left(\frac{g\mu_B B \tau}{\pi\hbar}\right) \\
	p_1 &=& 1-p_0 = \sin^2\left(\frac{g\mu_B B \tau}{\pi\hbar}\right).
\end{eqnarray}

During the measurement, we count the number of photons collected from the NV and assign the result to either \nvm or \nvz.  For the two spin states, we denote the probabilities of measuring \nvm as $\tilde{\beta}_{0}$ and $\tilde{\beta}_{1}$.  For a perfect charge state, $\tilde{\beta}_i = \beta_i$.  The signal $S$ that we record is the fraction of repetitions of the experiment for which the result is \nvm.  The expected signal for a measurement of the superposition state $|\psi(\tau)\rangle$ (eq.\ \ref{eq:SpinEchoState}) will be:
\begin{equation}
	S = p_0\tilde{\beta}_0 + p_1\tilde{\beta}_1.
\end{equation}
The minimum detectable change in the magnitude of the magnetic field, $\delta B$, is that which shifts the mean $\langle S \rangle$, by the width, $\sigma_S$:
\begin{equation}
	\delta B = \frac{\sigma_S}{\partial \langle S \rangle /\partial B} = \frac{\pi\hbar}{g\mu_B\tau} \frac{\sigma_S}{\tilde{\beta}_0 - \tilde{\beta}_1},
\end{equation}
where we have taken $p_0 = p_1 = 1/2$, to maximize the slope of the signal with respect to a change in magnetic field amplitude.  The sensitivity is related to $\delta B$ by the square root measurement time.  Thus:
\begin{equation}
	\label{eq:genericsensitivity}
	\eta = \delta B \sqrt{\tau + t_I + t_R} = \frac{\pi\hbar}{2g\mu_B} \times \sigma_R \times \sqrt{\frac{\tau + t_I + t_R}{\tau^2}}.
\end{equation}
where we have defined $\sigma_R = \frac{2g\mu_B\tau}{\pi\hbar}\frac{\sigma_S}{\partial S / \partial B}$ to be the readout noise per shot, normalized so that for a measurement where spin projection noise is the only source of uncertainty, $\sigma_R=1$.   In general, $\sigma_R$ will be a function of $t_I$ and $t_R$, so that improvements in $\sigma_R$ must be balanced with the associated requirements in overhead time.

We now derive an expression for $\sigma_R$ in the case of spin readout based on the SCC mechanism.  Recall that we consider operation at the point $p_0 = p_1 = 1/2$:
\begin{eqnarray}
	\sigma_R^{SCC} &=& \frac{2g\mu_B\tau}{\pi\hbar}\frac{\sigma_S}{\partial \langle S \rangle / \partial B} \\
	\langle S\rangle &=& p_0\beta_0 + (1-p_0)\beta_1 \\
	\frac{\partial \langle S\rangle}{\partial B} &=& \frac{g\mu_B\tau}{\pi\hbar}(\beta_0-\beta_1) \\
	\sigma_S^2 &=& \frac{\tilde{\beta}_0+\tilde{\beta}_1}{2} - \langle S\rangle^2 = \frac{1}{2}(\tilde{\beta}_0+\tilde{\beta}_1) - \frac{1}{4} (\tilde{\beta}_0+\tilde{\beta}_1)^2 \nonumber \\
	&=& \frac{1}{4}(\tilde{\beta}_0+\tilde{\beta}_1)(2-\tilde{\beta}_0-\tilde{\beta}_1) \\
	\sigma_R^{SCC} &=& \sqrt{\frac{(\tilde{\beta}_0+\tilde{\beta}_1)(2-\tilde{\beta}_0-\tilde{\beta}_1)}{(\tilde{\beta}_0-\tilde{\beta}_1)^2}}
\end{eqnarray}

\section{Measuring readout noise vs. readout time}

\begin{figure}
\centering
\includegraphics{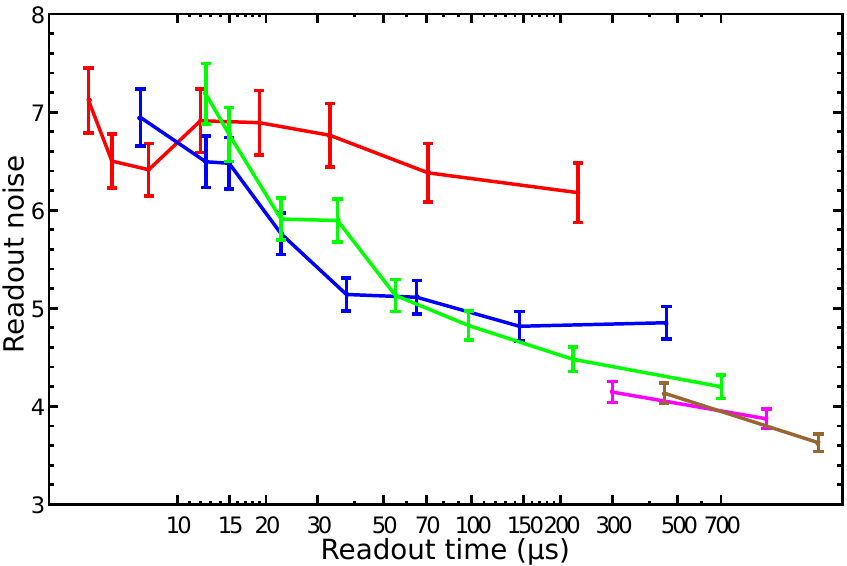}
\caption{Measured values of $\sigma_R^{SCC}$ for a set of powers and times chosen to optimize the readout noise for a photon threshold of 1 photon (red points), 2 photons (blue points), 3 photons (green points), 4 photons (magenta points), or 6 photons (brown points).}
\label{fig:SCC_SigmaRFull}
\end{figure}

We measured $\tilde{\beta}_{0,1}$ for a range of readout powers.  For each power, we measured the ionization and photon count rates, and optimized the readout time to minimize $\sigma_R^{SCC}$ for the set of threshold photon numbers $n_{thresh}=\{1,2,3,4,6\}$ (for $n_{thresh}=\{4,6\}$ we only used the two lowest powers).  For each (power, time, threshold) combination, we ran the SCC sequence after a fast initialization and iterated 100,000 times.  The results of all measurements are shown in Fig.~\ref{fig:SCC_SigmaRFull}.  As with the measurement of $\mathcal{F}_C(t_R)$, we used each photon number threshold in its optimal range for the plot in Fig.~4.

In order to estimate the magnetometer sensitivity from this measurement, we need an approximate functional dependence for $\sigma_R^{SCC}(t_R)$, valid over the measurement range.  We used a fit function of the form $f(t_R)=1+at_R^{-b}$, with fitting parameter values $a=7.54$, $b=0.146$.  The projected $\eta$ for a given $\tau$ is found by minimizing $f(t_R)\sqrt{t_I+t_R+\tau}$ with respect to $t_R$.

\section{Spin polarization}

\begin{figure}
\centering
\includegraphics{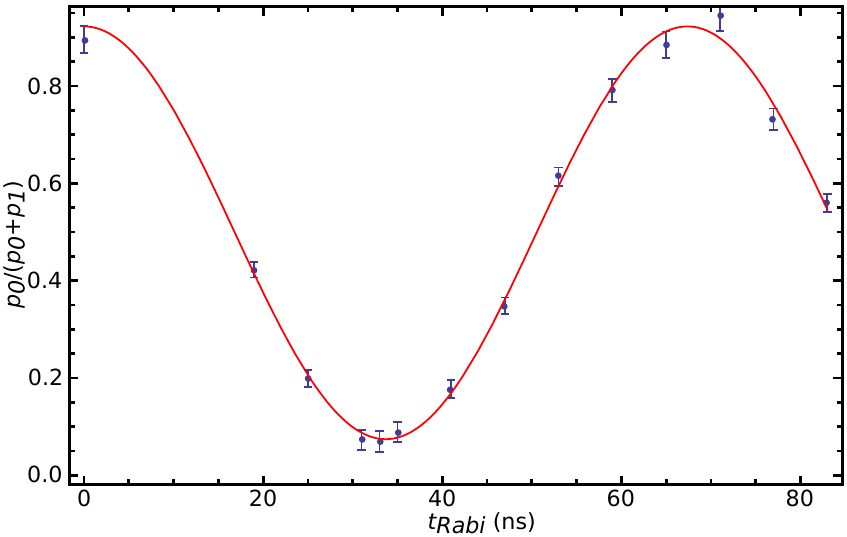}
\caption{Polarization measurement.  The NV is initialized with a pump-probe sequence (\SI{150}{\ns}, \SI{300}{\micro\watt}, 532-nm pump pulse followed by \SI{900}{\ns}, \SI{11}{\micro\watt} 594-nm probe pulse.  Microwaves of duration $t_{Rabi}$ rotate the spin into a superposition of \msz and \mso, and the fluorescence decay is recorded upon excitation with a \SI{50}{\pico\second} pulsed source.  From the decay, the population in \msz is extracted (blue points) and fitted by a sinusoid (red curve) to find the initial spin polarization.}
\label{fig:SCC_Polarization}
\end{figure}

The measurements of $\sigma_R^{SCC}$ all include the effect of imperfect spin polarization.  Upon excitation, an initial \nvm state of \mso is excited and decays preferentially into the singlet state, where it can decay back to \msz or \mso.  The limited branching ratios at each decay step result in imperfect spin polarization.  To measure the electron spin polarization of the NV, we perform measurement of the triplet excited state lifetime as the NV undergoes Rabi nutations in the ground state\cite{1367-2630-13-2-025013}.  We use the pump-probe sequence to initialize the center into \nvm, and apply a microwave pulse of duration $t_{Rabi}$ followed by a \SI{50}{\pico\second} pulse of 532-nm light.  The fluorescence intensity $I(t)$ subsequent to the \SI{50}{\pico\second} pulse are recorded with a time-correlated single photon counting module (PicoHarp 300, PicoQuant) and conditioned on detection of a probe photon.  We fit the fluorescence decay by a sum of two exponentially modified Gaussian distributions:
\begin{equation}
I(t) = p_0(t_{Rabi}) F(t,\tau_0) + p_1(t_{Rabi}) F(t,\tau_1) + c.
\end{equation}
Here, $p_0$ and $p_1$ are the amplitude in \msz and \mso, $t$ is the delay after the \SI{50}{\pico\second} pulse, and $F(t,\tau)$ is an exponentially modified Gaussian with exponential decay constant $\tau$.  The decay constants were found to be $\tau_0 = \SI{18.2}{\nano\second}$ and $\tau_1 = \SI{7.9}{\nano\second}$.  We find the fraction of population in \msz for each value of $t_{Rabi}$ (blue points in Fig.~\ref{fig:SCC_Polarization}) and fit the result with the function:
\begin{equation}
	p_{0}(t_{Rabi}) = a\cos(\omega t_{Rabi}) + c,
\end{equation}
with $a=0.42\pm0.01$ and $c=0.50\pm0.01$, so that the initial polarization is $p_{0}(0) = 0.92\pm0.01$.

\end{document}